\documentclass[a4paper,11pt,headings=big,DIV=12]{scrartcl}
\usepackage{amsmath,amssymb}
\usepackage{color,xcolor}
\definecolor{blue}{rgb}{0,0,0.5}
\usepackage[colorlinks,linkcolor=blue,urlcolor=blue,citecolor=blue]{hyperref}
\usepackage{graphicx}
\addtokomafont{disposition}{\rmfamily\boldmath}
\usepackage{cite}
\usepackage{xspace}
\usepackage{relsize}
\usepackage{bm}
\usepackage[utf8]{inputenc}

\newcommand{\HEPfit}{\texttt{HEPfit}\xspace}
\newcommand{\flavio}{\texttt{flavio}\xspace}
\allowdisplaybreaks

\begin{document}

\begin{center}
{\LARGE\bfseries \boldmath
\vspace*{1.5cm}
Constraints on new physics from radiative $B$ decays
}\\[0.8 cm]
{\Large%
Ayan Paul$^a$ and David M. Straub$^b$
\\[0.5 cm]
\small
$^a$\,INFN, Sezione di Roma, I-00185 Rome, Italy
\\[0.2cm]
$^b$\,Excellence Cluster Universe, TUM, Boltzmannstr.~2, 85748~Garching, Germany
}
\\[0.5 cm]
\small
E-Mail:
\texttt{\href{mailto:apaul2@alumni.nd.edu}{apaul2@alumni.nd.edu}, \href{mailto:david.straub@tum.de}{david.straub@tum.de}}
\\[0.2 cm]
ORCID:
\texttt{\href{https://orcid.org/0000-0002-2156-4062}{0000-0002-2156-4062}, \href{http://orcid.org/0000-0001-5762-7339}{0000-0001-5762-7339}}
\end{center}

\bigskip

\begin{abstract}\noindent
A new phase for the measurements of radiative decay modes in $b \to s$ transitions has started with new measurements of exclusive modes by LHCb
and with Belle-II showing distinctive promises in both inclusive and exclusive channels.
After critically reviewing the hadronic uncertainties in exclusive radiative decays,
we analyze the impact of recent measurements of the
branching ratio and mass-eigenstate rate asymmetry in $B_s\to\phi\gamma$ and
of the angular distribution of $B\to  K^*e^+e^-$ at low  $q^2$
on new physics in the $b\to s\gamma$ transition.
\end{abstract}

\setcounter{tocdepth}{2}
\tableofcontents

\newpage

\section{Introduction}

The radiative flavour-changing neutral current (FCNC) transition $b\to s\gamma$
is a crucial probe of physics beyond the Standard Model (SM). Its strong 
sensitivity to new physics (NP) stems from the helicity flip required for the
dipole transition, in addition to the CKM and loop suppression. Beyond the SM,
new sources of chirality breaking -- like the trilinear couplings in the MSSM
or masses of heavy vector-like quarks -- can strongly modify the $b\to s\gamma$
transition or can enhance the amplitude with a right-hand polarized photon,
$b_L\to s_R\gamma_R$, that is suppressed by a factor $m_s/m_b$ at leading order
in the SM with respect to $b_R\to s_L\gamma_L$.

The strongest constraint on the $b\to s\gamma$ transition comes from the
measurement of the branching ratio of the inclusive decay $B\to X_s\gamma$,
which is theoretically particularly clean as the decay rate is given simply
by the quark-level decay rate in the heavy $b$ quark limit.
The branching ratios of the exclusive channels $B^{0,+}\to K^*\gamma$
have been measured even more precisely, but their theoretical estimations are afflicted with considerable
hadronic uncertainties.
These uncertainties also affect the branching ratio of the exclusive decay
$B_s\to \phi\gamma$, that has been measured recently.
But branching ratios alone cannot distinguish between the different
photon helicities.
The most promising observables to make this distinction
are ones that vanish in the limit of purely
left-handed photon polarization. A prominent example is the
mixing-induced CP asymmetry in $B^0\to K^*\gamma$
\cite{Atwood:1997zr, Ball:2006cva}.
The angular distribution of the exclusive decay $B\to K^*(\to K\pi)e^+e^-$
at very low invariant mass of the dielectron pair also offers observables
that have this property \cite{Kruger:2005ep}. Very recently, a previously unmeasured observable of this type has been
measured by LHCb \cite{Aaij:2016ofv}: the mass-eigenstate rate asymmetry $A_{\Delta\Gamma}$ in
$B_s\to \phi\gamma$, that quantifies the decay rate difference of the two
$B_s$ mass eigenstates into the $\phi\gamma$ final state \cite{Muheim:2008vu}.
All these observables have a complementary dependence on the CP phases and
chirality of the new physics contributions and thus lead to powerful constraints
on physics beyond the SM, when considered together.
Fits of new physics in the $b\to s\gamma$ transition have been performed
in the past in the context of global $b\to s$ fits (see e.g.\
\cite{Beaujean:2013soa,Altmannshofer:2014rta,Descotes-Genon:2015uva,Hurth:2016fbr}).
Dedicated fits to radiative decays have also been performed
\cite{DescotesGenon:2011yn,Becirevic:2012dx}; these
have the advantage that they are less dependent on theory uncertainties or
potential NP effects in observables that are less sensitive to the radiative
dipole transition. The main novelty of our analysis with respect to the latter
analyses is that we are including the new measurements of the $B\to K^*e^+e^-$
angular observables at low $q^2$ and the $B_s\to\phi\gamma$ mass-eigenstate rate
asymmetry for the first time. We also emphasize the importance of the direct
CP asymmetry in $B\to K^*\gamma$ as a constraint on new physics.
Furthermore, our numerics relies completely on
open source codes, which makes it possible for the interested reader to reproduce our numerical
predictions and plots and to study the impact of modifying any parametrical
assumptions made.

The rest of the paper is organized as follows.
In section~\ref{sec:obs}, we
define the effective Hamiltonian and discuss the observables in inclusive and
exclusive decays that are sensitive to NP. We put a particular emphasis on
the theory uncertainties that affect the exclusive decays.
In section~\ref{sec:num}, we present a numerical analysis of the current
constraints on new physics in the $b\to s\gamma$ transition, taking into
account the theoretical uncertainties.
Section~\ref{sec:conclusions} contains our conclusions.

\section{Observables}\label{sec:obs}
In the following sections we will define our operator basis and then move on to defining the inclusive observables that we use in this work. For discussing the exclusive observables it is important that we include a short discussion on our choice of form factors and how we treat all the hadronic uncertainties that are inherent to these observables. The subtleties of oscillations also appear in some of them and are hence discussed too.

\subsection{Effective Hamiltonian}

The effective Hamiltonian relevant for inclusive and exclusive decays based
on the $b \to s\gamma$ transition in the SM and beyond can be written as%
\footnote{For simplicity, we have omitted from \eqref{eq:Heff} -- but not
from our numerics -- electroweak penguin
operators and doubly Cabibbo-suppressed terms.
We also neglect new physics in four-quark operators.
}
\begin{equation}
\mathcal H_\text{eff} = -\frac{4G_F}{\sqrt{2}} \lambda_t \left(\sum_{i=1}^8  C_i Q_i + \sum_{i=7}^8  C_i' Q_i' \right) 
\label{eq:Heff}
\end{equation}
where $\lambda_i=V_{ib}V_{is}^*$
and $Q_\text{1--6}$ are the SM four-quark operators (see e.g. \cite{Blake:2016olu}	
for explicit expressions) and $Q_7^{(\prime)}$ and $Q_8^{(\prime)}$, the
electromagnetic and chromomagnetic dipole operators, are given by
\begin{align}
Q_7^{(\prime)} &= \frac{e}{16\pi^2}m_b (\bar s_{L(R)}\sigma_{\mu\nu}b_{R(L)}) F^{\mu\nu} \,,
&
Q_8^{(\prime)} &= \frac{g_s}{16\pi^2}m_b (\bar s_{L(R)}\sigma_{\mu\nu}T^a b_{R(L)}) G^{a\mu\nu} \,.
\end{align}
For the Wilson coefficient $C_7$, it is customary to define a regularization
scheme independent ``effective'' coefficient
\begin{equation}
 C_7^\text{eff}(\mu) = C_7(\mu) + \sum_{i=1}^6 y_i C_i(\mu)
\end{equation}
where $y=(0,0,-\frac{1}{3},-\frac{4}{9},-\frac{20}{3},-\frac{80}{9})$ in a scheme
with fully anti-commuting $\gamma_5$ and in the operator basis used in
\cite{Chetyrkin:1996vx}.
Its numerical value at the scale $\mu=4.8\,\text{GeV}$ is  $C_7^\text{eff}=-0.2915$
\cite{Blake:2016olu}, including next-to-next-to-leading order QCD and
next-to-leading order electroweak corrections\footnote{%
See \cite{Buras:2011we} for an account of the contributions to this
calculation.}, with an uncertainty at the per
mille level. The chirality-flipped coefficient $C_7'$ is given by
$C_7'=\frac{m_s}{m_b}C_7$ in the SM.

\subsection{Inclusive radiative decay}

The branching ratio of the inclusive decay $B\to X_s\gamma$ can be predicted
to a high accuracy, as the width is given by the width of the quark-level
decay $b\to s\gamma$ up to corrections that vanish in the infinite $b$ quark
mass limit. It is to be noted that experiments only resolve
photons down to a minimum energy $E_0$. At present, the best compromise between
experimental and theoretical sensitivity is given by $E_0=1.6\,\text{GeV}$.
To further reduce theoretical uncertainties, the rate can be normalized to the
precisely measured semi-leptonic inclusive decay $B\to X_c\ell\nu$, assuming
it to be unaffected by NP. The resulting prediction reads
\begin{equation}
\text{BR}(B\to X_s\gamma)_{E_\gamma>E_0} =
\text{BR}(B\to X_c\ell\nu)  \left| \frac{\lambda_t}{V_{cb}} \right|^2 \frac{6\alpha_\text{em}}{\pi C}
\left[ P(E_0) + \delta_\text{nonp.}\right]
\end{equation}
where $\delta_\text{nonp.}$ is a non-perturbative contribution that is estimated
at approximately 5\% of the SM branching ratio \cite{Benzke:2010js}.
At leading order, the function $P(E_0)$ is given by
\begin{equation}
P(E_0)_\text{LO} = |C_7^\text{eff}|^2 + |C_7'|^2 \,.
\end{equation}
Within the SM, corrections of order $\alpha_s^2$ and
$(\lambda_u/\lambda_t)^2$ have been computed (see \cite{Misiak:2015xwa} for an account
on the state of the art).

Apart from the (CP-averaged) branching ratio, also the
direct CP asymmetry in the inclusive decay can be measured and used in principle
to constrain CP-violating NP contributions. While the direct CP asymmetry in
$B\to X_s\gamma$ is plagued by poorly known long-distance contributions
\cite{Benzke:2010tq}, the direct CP asymmetry in the untagged decay
$B\to X_{s+d}\gamma$ is free from such contributions and thus theoretically
clean. Unless the NP contributions to the Wilson coefficients in the
$b\to d\gamma$ transition are much larger than their counterparts in
$b\to s\gamma$, the NP contribution to the untagged CP asymmetry is dominated by
the $b\to s\gamma$ Wilson coefficients (see e.g. \cite{Hurth:2003dk}). 
However, it turns out that even in the presence of NP, the CP asymmetry is
numerically small since the strong phases, generated first at NLO, are small.
For instance, we find that even for $\text{Im}(C_7^\text{NP})=0.5$, the untagged CP
asymmetry stays below 1\%, to be compared to the current world average
of $3.2\pm 3.4\%$ \cite{Amhis:2014hma}. We conclude that the inclusive CP asymmetry currently does
not constitute a relevant constraint on new physics.

\subsection{Exclusive radiative decays}

While inclusive decay modes are theoretically cleaner and hence more accurately
predictable, exclusive modes are more easily measured in experiments because of
their distinctive final states. However, the
observables in these exclusive modes are usually shrouded in hadronic
uncertainties that are difficult to compute rendering them less conducive for
prediction within a theoretical model, be it the SM or a model of new physics.
These uncertainties are twofold: on one hand, predicting the observables
requires the knowledge of the hadronic form factors of the relevant
$B\to V$ transition. On the other hand, the radiative decays receive
contributions where the photon does not participate in the -- purely
hadronic -- hard interaction, a contribution which cannot be expressed in terms of form
factors. For predicting these ``non-factorizable'' contributions,
different approaches have been advocated.
In the following, we give a general parametrization of these contributions
and try to estimate their size and uncertainty, based on existing calculations
and estimates.

\subsubsection{General parametrization of amplitudes}\label{sec:amps}

The polarized amplitudes of a general $B_q\to V\gamma$ decay, where
$q=u$, $d$, or $s$, and
$V$ is a vector meson,
can be written as

\begin{align}
\mathcal M(\bar B_q\to \bar V\gamma_{L}) &= N\left(C_7^\text{eff} \, T_1(0) + h_L \right)S_L
\\
\mathcal M(\bar B_q\to \bar V\gamma_{R}) &= N\left(C_7' \, T_1(0) + h_R \right)S_R
\\
\mathcal M( B_q\to  V\gamma_{L}) &= N^* \left(C_7^{\prime *} \, T_1(0) + h_R \right)S_L
\\
\mathcal M( B_q\to  V\gamma_{R}) &= N^*\left(C_7^{\text{eff}*} \, T_1(0) + h_L \right)S_R
\end{align}
where
\begin{equation}
N = -\frac{4 G_F }{\sqrt{2}}\lambda_t \frac{e}{8\pi^2} \,,
\end{equation}
\begin{equation}
 S_{L,R} =
 \epsilon^{\mu\nu\rho\sigma} e_\mu^*\eta_\nu^*p_\rho q_\sigma
 \pm i
 \left[
 (e^*\eta^*)(pq)-(e^*p)(\eta^*q)\right],
\end{equation}
$T_1(0)$ is the process-dependent $B_q\to V$ tensor form factor at $q^2=0$,
and we have neglected doubly Cabibbo-suppressed terms of order
$\lambda_u/\lambda_t$. The process-dependent\footnote{We omit labels on
process-dependent quantities like the form factors or $h_{L,R}$
in the following to avoid clutter.} quantities $h_L$ and
$h_R$ denote contributions
of the weak hadronic Hamiltonian, i.e. from the operators $Q_\text{1--6}$
and $Q_8$.
In general, they are complex quantities but enter the amplitudes without
a conjugate as their phases are CP-even.

Equivalently, we can express the hadronic contributions as 
process-dependent shifts of the
Wilson coefficients, 
\begin{align}
\mathcal M(\bar B_q\to \bar V\gamma_{L(R)}) &= \left(N \, \,\mathcal C_7^{(\prime)}\, T_1(0) \right)S_{L(R)}
\,,&
\mathcal M( B_q\to  V\gamma_{R(L)}) &= \left(N^* \, \,\overline{\mathcal C}_7^{(\prime)}\, T_1(0) \right)S_{R(L)}
\,,
\end{align}
where we have defined
\begin{align}
\mathcal C_7 &=  C_7^\text{eff} + \Delta C_7 = C_7^\text{eff} + \frac{h_L}{T_1(0)}
\\
\overline{\mathcal C}_7 &=   C_7^{\text{eff} *} + \Delta C_7 = C_7^{\text{eff} *} + \frac{h_L}{T_1(0)}
\end{align}
and analogously for the primed coefficient.

\subsubsection{Discussion of form factor uncertainties}\label{sec:ff}

The main source of uncertainty in the branching ratios stems from
the form factor $T_1(0)$. In our numerical analysis,
we use a recent update \cite{Straub:2015ica} of a light-cone sum rules calculation
\cite{Ball:2004rg} of the full QCD form  factors yielding
\begin{subequations}
\label{eq:T1}
\begin{align}
\label{eq:T1Ks}
T_1(0) &= 0.282 \pm 0.031 &&\text{for } B\to K^*\gamma
\,,\\
\label{eq:T1phi}
T_1(0) &= 0.309 \pm 0.027 &&\text{for } B_s\to \phi\gamma
\,,
\end{align}
\end{subequations}
where the form factor is defined at the $\overline{\text{MS}}$ scale
$\mu_b=4.8\,\text{GeV}$.
A combined fit of the LCSR results and a recent lattice computation valid
at high $q^2$ \cite{Horgan:2015vla}
yields
\begin{subequations}
\label{eq:T1C}
\begin{align}
\label{eq:T1KsC}
T_1(0) &= 0.312 \pm 0.027 &&\text{for } B\to K^*\gamma
\,,\\
\label{eq:T1phiC}
T_1(0) &= 0.299 \pm 0.012 &&\text{for } B_s\to \phi\gamma
\,.
\end{align}
\end{subequations}
To be conservative, we will stick to the LCSR result in the following.

\subsubsection{Discussion of hadronic contributions}\label{sec:hadronic}

Concerning the hadronic effects encoded in $\Delta C_7$,
we include the following known contributions:
\begin{itemize}
 \item $O(\alpha_s)$ vertex corrections involving matrix elements of the
 current-current operators $Q_\text{1,2}$~\cite{Asatryan:2001zw};
 \item hard spectator scattering involving matrix elements of
 $Q_\text{1--6}$ and $Q_8$ at leading order in $\Lambda/m_b$ from QCD
 factorization \cite{Beneke:2001at};
 \item weak annihilation at $O(\Lambda/m_b)$ from QCD factorization \cite{Kagan:2001zk,Feldmann:2002iw,Beneke:2004dp}.
\end{itemize}
The numerical central values of all the contributions are shown in
table~\ref{tab:deltaC7} for the three relevant transitions.
The vertex correction is by far the largest contribution.
At $O(\alpha_s)$, it suffers from a sizable uncertainty due to the dependence
on the charm quark mass scheme. This problem is known from the inclusive decay
(see e.g. \cite{Gambino:2001ew,Asatrian:2005pm})
and could be resolved by including $O(\alpha_s)^2$ corrections.
Importantly, since the vertex correction
factorizes into a product of the universal part and a $B\to V$ form factor, its
contribution to the shift $\Delta C_7$ is the same for all exclusive decays
based on the $b\to s$ transition.
Spectator scattering turns out to be independent of the spectator quark charge
at leading order in QCDF,  and is, thus, also universal for all three
transitions, up to small parametric differences.
Our numerical results for the vertex correction and spectator scattering
agree well with an alternative derivation using SCET \cite{Ali:2007sj}.
The weak annihilation contribution is proportional to the spectator
quark charge at the order considered, so it differs by a factor $-2$ between
the charged and neutral $B\to K^*\gamma$ transitions. For $B_s\to\phi\gamma$,
there is an additional penguin contribution where the $b\to s$ transition connects
the initial $b$ quark with the spectator $s$ quark, but it is suppressed by
small Wilson coefficients, so the differences with respect to $B^0\to K^*\gamma$
remain small.

The following contributions are instead \textit{not} included in our central values for
$\Delta C_7$:
\begin{itemize}
 \item QCDF power corrections to spectator scattering involving
 $Q_8$, which are endpoint divergent \cite{Kagan:2001zk,Feldmann:2002iw,Beneke:2004dp};
 \item Contributions to weak annihilation and spectator scattering involing $Q_8$
 beyond QCDF that have been computed with light-cone sum rules (LCSR)
\cite{Ball:2006eu,Dimou:2012un,Lyon:2013gba};
 \item Soft gluon corrections to quark loop spectator scattering, in particular
 the charm loop that comes with a large Wilson coefficient
 \cite{Ball:2006eu,Khodjamirian:2010vf,Muheim:2008vu}.
\end{itemize}
Of these contributions, the soft gluon correction to the charm loop is expected
to be numerically dominant in view of the large Wilson coefficient $C_2$.
This effect has been estimated for $B\to K^*\gamma$
in \cite{Ball:2006eu} using LCSR and finding, in our conventions,
\begin{align}
\Delta C_7|_\text{soft} &= (-0.52 \pm 0.43) \times 10^{-2}
\,, &
\Delta C_7'|_\text{soft} &= (0.13 \pm 0.20) \times 10^{-2}
\,.
\end{align}
This calculation was refined  and applied to $B_s\to\phi\gamma$
in \cite{Muheim:2008vu}, finding
\begin{align}
\Delta C_7|_\text{soft} &= (1.11 \pm 0.78) \times 10^{-2} \,e^{i(255\pm15^\circ)}
\,, &
\Delta C_7'|_\text{soft} &= (0.42 \pm 0.29)\times 10^{-2} \,e^{i(106\pm15^\circ)}
\,.
\label{eq:DC7-Zwicky}
\end{align}
A different approach, also using LCSR, was taken in \cite{Khodjamirian:2010vf}
finding\footnote{The authors of \cite{Khodjamirian:2010vf} confirmed to us
an erroneous sign in their eq.~(6.3).}
\begin{align}
\Delta C_7|_\text{soft} &= (+1.2^{+0.9}_{-1.6})\times 10^{-2}
\,, &
\Delta C_7'|_\text{soft} &\approx 0
\,.
\end{align}
Given the marginal agreement between the different estimates,
we omit this effect and account for its omission -- and all remaining neglected
effetcts -- by adding
an uncertainty on the real and imaginary parts of $\Delta C_7$ of
$\pm 1.5\times10^{-2}$, which includes the central values of all three estimates
and is much larger than the difference between the
QCDF and LCSR computations of weak annhilation and spectator scattering.

Concerning $\Delta C_7'$, i.e.\ hadronic contributions to the ``wrong-helicity''
amplitude, we do not attempt a prediction but only parametrize our ignorance.
While simple power counting leads to a relatively large possible range for this quantity \cite{Grinstein:2004uu,Grinstein:2005nu}, based on more refined parametric
arguments implying the existence of a helicity suppression \cite{Becirevic:2012dx,Jager:2012uw,Jager:2014rwa} as well as on the above
numerical estimates of the soft gluon charm loop contribution to $\Delta C_7'$
from LCSR,
we assume an uncertainty of $\pm0.4\times10^{-2}$ on the real and imaginary parts
of $\Delta C_7'$ and consider this to be a conservative choice.

\begin{table}[tbp]
\renewcommand{\arraystretch}{1.2}
\centering
\begin{tabular}{ccccc}
\hline
& $B^0\to K^*\gamma$ & $B^+\to K^*\gamma$ & $B_s\to \phi\gamma$ \\
\hline
Vertex corrections & \multicolumn{3}{c}{$-(7.8\pm1.0)-(1.1\pm0.3)i$} \\
Spectator scattering $Q_\text{1--6}$ & $-0.7-1.3i$ & $-0.7-1.3i$ & $-0.7-1.7i$ \\
Spectator scattering $Q_\text{8}$ & $-0.3$ & $-0.3$ & $-0.4$\\
Weak annihilation & $-0.4$ & $+0.9$ & $-0.5$ \\
\hline
\end{tabular}
\caption{Contributions to $\Delta C_7$ in units of $10^{-2}$ for the three decays.
We only show the uncertainties for the (numerically dominant) vertex corrections
that are process independent.}
\label{tab:deltaC7}
\end{table}

\subsubsection{Exclusive radiative observables}\label{sec:ex}

With the notation introduced in section~\ref{sec:amps}, the helicity-summed
and CP-averaged branching ratio can be written as
\begin{equation}
\text{BR}(B_q\to V\gamma) 
=
\tau_{B_q}\frac{G_F^2\alpha_\text{em}m_{B_q}^3 m_b^2}{32\pi^3} \left(1-\frac{m_V^2}{m_B^2}\right)^3 |\lambda^t|^2
\left(|\mathcal C_7|^2 +  |\mathcal C_7'|^2 \right) T_1(0)\,.
\label{eq:BRBVgamma}
\end{equation}
where the quantities $\alpha_\text{em}$, $m_b$, $\mathcal C_7^{(\prime)}$, and
$T_1(0)$ are to be understood as $\overline{\text{MS}}$ quantities at the scale
$\mu_b$ that we take to be $4.8\,\text{GeV}$.

In the decay of the neutral mesons $B_d$ and $B_s$ to CP eigenstates,
like $B_s\to\phi\gamma$ or $B^0\to K^*(\to K_S\pi^0)\gamma$,
the time-dependent CP
asymmetry leads to additional observables, thanks to meson-antimeson mixing.
It reads
\begin{align}
A_\text{CP}(B_q(t)\to V\gamma)
&=
\frac
{\Gamma(\bar B_q(t)\to  \bar V\gamma)-\Gamma(  B_q(t)\to   V\gamma)}
{\Gamma(\bar B_q(t)\to  \bar V\gamma)+\Gamma(  B_q(t)\to   V\gamma)}
\\
&=
\frac
{S(B_q\to V\gamma) \sin(\Delta M_q t)+A_\text{CP}(B_q\to V\gamma) \cos(\Delta M_q t)}
{\cosh(y_q t/\tau_{B_q}) -A_{\Delta\Gamma}(B_q\to V\gamma) \sinh(y_q t/\tau_{B_q})},
\end{align}
containing the direct CP asymmetry $A_\text{CP}$, the mixing-induced
CP asymmetry $S$ and the mass-eigenstate rate asymmetry $A_{\Delta\Gamma}$.

In the case of the $B^0\to K^*\gamma$, the direct CP asymmetry $A_\text{CP}$ also
coincides\footnote{This fact relies on $\Delta M_d\ll\Gamma_d$ and thus does not
hold in the $B_s$ system.} to excellent approximation with the \textit{time-integrated}
CP asymmetry, while $A_{\Delta\Gamma}$ can be neglected, since
the normalized width difference
\begin{equation}
y_q = \Delta \Gamma_q/(2\Gamma_q)=\tau_{B_q}\Delta\Gamma_q/2 
\end{equation}
is tiny for $B_d$ mesons.
In terms of the quantities defined above, one can write
\begin{equation}
A_\text{CP}(B^0\to K^*\gamma) \times \text{BR}(B^0\to K^*\gamma)
= 2 \, \text{Im}\,C_7^\text{eff} ~ \text{Im}\,\Delta C_7
+ 2 \, \text{Im}\,C_7' ~ \text{Im}\,\Delta C_7' +\ldots
\label{eq:ACPKstar}
\end{equation}
where the ellipsis includes in particular the doubly Cabibbo-suppressed SM
contribution.
Since $\Delta C_7'$ is expected to be very small, $A_\text{CP}$ mostly constrains
the imaginary part of the NP contribution to $C_7$ \cite{Altmannshofer:2014rta}.
As seen from table~\ref{tab:deltaC7}, spectator scattering induces a 
non-negligible strong phase to $\Delta C_7$, which in combination with the
precise measurements of $A_\text{CP}$ -- that actually do not require a time-dependent
analysis -- make it a strong constraint as we will
discuss in section~\ref{sec:num}.

The quantity $A_{\Delta\Gamma}$
is only relevant for the $B_s$ decay.
Due to the sizable width difference $y_s\approx 6\%$ in the $B_s$ system,
$A_{\Delta\Gamma}$ can be extracted from the \textit{untagged}
time-dependent decay rate of $B_s\to \phi\gamma$, 
\begin{align}
\text{BR}(B_s(t)\to \phi\gamma) =
\text{BR}(B_s\to \phi\gamma)\,e^{-t/\tau_{B_s}}\left[
\cosh\!\left(\frac{y_s t}{\tau_{B_s}}\right)
-A_{\Delta\Gamma}(B_s\to \phi\gamma) \sinh\!\left(\frac{y_s t}{\tau_{B_s}}\right)
\right].
\end{align}
This also implies that the
time-integrated branching ratio of the $B_s\to\phi\gamma$ decay measured
experimentally does not coincide with the theoretical branching ratio in the
absence of $B_s$-$\bar B_s$ mixing given in \eqref{eq:BRBVgamma}. The two
quantities are related as
\begin{align}
\overline{\text{BR}}(B_s\to \phi\gamma)=
\left[\frac{1-A_{\Delta\Gamma}(B_s\to \phi\gamma)\, y_s}{1-y_s^2}\right]
\text{BR}(B_s\to \phi\gamma).
\end{align}

Particularly simple expressions for $S$ and $A_{\Delta\Gamma}$  can be obtained
in the approximation $|\mathcal C_7^{(\prime)}|\approx|\overline{\mathcal C}_7^{(\prime)}|$
(the violation of this relation generates $A_\text{CP}$).
Then one can write
$\mathcal C_7=|\mathcal C_7|e^{i\phi_7}e^{i\delta_7}$,
$|\overline{\mathcal C}_7|=|\mathcal C_7|e^{-i\phi_7}e^{i\delta_7}$,
and analogously for the primed coefficients, leading to
\begin{align}
\label{eq:Sphigamma}
S(B_s\to\phi\gamma) &=
\sin(2\chi)
\,\sin(\phi_7+\phi_7' - \phi_s^\Delta)
\cos(\delta_7-\delta_7')
\,,
\\
\label{eq:ADeltaGamma}
A_{\Delta\Gamma}(B_s\to\phi\gamma) &=
\sin(2\chi)
\,\cos(\phi_7+\phi_7' - \phi_s^\Delta)
\cos(\delta_7-\delta_7')
\,,
\end{align}
where we have introduced
\begin{equation}
\tan\chi \equiv \left|\frac{{\mathcal C}_7'}{{\mathcal C}_7}\right|
\label{eq:chi}
\end{equation}
and $\phi_s^\Delta$ is a NP contribution to the $B_s$ mixing phase.
Analogously, the mixing-induced CP asymmetry in $B^0\to K^*(\to K_S\pi^0)\gamma$ is given by 
\begin{equation}
\label{eq:SKsgamma}
S(B^0\to K^*\gamma) =
\sin(2\chi)
\,\sin(\phi_7+\phi_7' - 2\beta - \phi_d^\Delta - 2|\beta_s|)
\cos(\delta_7-\delta_7')
\,.
\end{equation}
where $\beta = \arg(V_{tb}^*V_{td})$, $\beta_s = \arg(V_{tb}^*V_{ts})$.
While we do not use these approximate expressions in our numerics, they
demonstrate clearly that
\begin{itemize}
\item both $S(B^0\to K^*\gamma)$ and $A_{\Delta\Gamma}(B_s\to\phi\gamma)$
measure the ratio of the amplitudes with right- and left-handed photons,
\item in the case of NP contributions leading to a sizable value for these
observables, the dependence on the strong phases is small, as they only enter
in the cosine terms.
\end{itemize}
The latter is in constrast to the direct CP asymmetry in
$B\to K^*\gamma$ which is instead strongly dependent on the strong phases
as seen in \eqref{eq:ACPKstar}.

\subsection{Exclusive semi-leptonic angular observables}

While strictly speaking $B^0\to K^{*0}(\to K\pi)e^+e^-$ is not a radiative decay, the angular analysis of
this rare semi-leptonic decay at a very low
dilepton invariant mass squared can give complementary constraints on the
Wilson coefficients $C_7$ and $C_7'$. This is true in particular for the 
observables \cite{Kruger:2005ep, Bobeth:2008ij, Altmannshofer:2008dz, Matias:2012xw}
\begin{align}
P_1&=A_T^{(2)}=\frac{S_3}{2S_2^s} \,,
A_T^\text{(Im)}=\frac{A_9}{2S_2^s} \,,
\end{align}
where we use the conventions used by the LHCb collaboration (see e.g.\
\cite{Blake:2016olu} for a dictionary between different conventions). $P_1$ is a CP-averaged
angular observable and $A_T^\text{(Im)}$ a T-odd CP asymmetry. Both observables
are quasi null tests of the SM. Importantly, both only depend on the
\textit{transverse} $B\to K^*e^+e^-$ helicity amplitudes. These helicity amplitudes 
also receive contributions from the weak hadronic Hamiltonian, but for $q^2\to0$
they coincide with the contributions to the $B^0\to K^*\gamma$ amplitudes.
Using again the approximation $|\mathcal C_7^{(\prime)}|\approx|\overline{\mathcal C}_7^{(\prime)}|$,
one can thus write (cf. \cite{Becirevic:2012dx})
\begin{align}
\lim_{q^2\to0}P_1  &=
\sin(2\chi)
\,\cos(\phi_7-\phi_7')\cos(\delta_7-\delta_7')
\,,
\label{eq:P1}
\\
\lim_{q^2\to0}A_T^\text{(Im)}  &= 
\sin(2\chi)
\,\sin(\phi_7-\phi_7')\cos(\delta_7-\delta_7')
\,.
\label{eq:ATIm}
\end{align}
where $\chi$ has been defined in \eqref{eq:chi}.

In practice, the observables are measured in finite $q^2$ bins,
as dictated by the experimental resolution and statistical precision.
While \eqref{eq:P1} and \eqref{eq:ATIm} no longer hold exactly in this case,
the impact of Wilson coefficients other than the ones contributing to
radiative decays is marginal even in the presence of NP, taking into account
other constraints.

Interestingly, the expressions \eqref{eq:P1} and \eqref{eq:ATIm} are very similar
to the expressions 
(\ref{eq:ADeltaGamma}) and (\ref{eq:Sphigamma}) for $A_{\Delta\Gamma}$ and
$S(B_s\to\phi\gamma)$, but there is an important difference: while the latter
depend on the sum of the weak phases of $\mathcal C_7$ and $\mathcal C_7'$,
the former depend on their difference. In scenarios with complex contributions
to both Wilson coefficients, they are thus complementary. Moreover, the
semi-leptonic angular observables are not affected by NP in meson-antimeson
mixing.

Before proceding to the numerical analysis, a comment is in order about the 
processes not discussed so far. While \textit{baryonic} exclusive radiative
decays such as $\Lambda_b\to \Lambda\gamma$ are promising but still awaiting
to be measured, LHCb has recently measured \cite{Aaij:2014wgo} a triple product asymmetry
in $B\to K_1(\to K\pi\pi)\gamma$ that is sensitive to the photon polarization
\cite{Gronau:2001ng,Kou:2010kn,Kou:2016iau}.
However at present this observable does not constitute a relevant constraint
on NP as it is plagued by large hadronic uncertainties.
In our notation, this asymmetry is proportional to $\cos(2\chi)$,
so it is less sensitive to small $\chi$ than the observables discussed above.

\begin{table}[tbp]
\renewcommand{\arraystretch}{1.2}
\centering
\begin{tabular}{ccccc}
\hline
Observable & SM prediction && Measurement & \\
\hline
$10^{4}\times\text{BR}(B\to X_s\gamma)_{E_\gamma>1.6\,\text{GeV}}$ & $3.36\pm 0.23$ &\cite{Misiak:2015xwa} & $3.27 \pm 0.14$ & \cite{Misiak:2017bgg} \\
$10^{5}\times\text{BR}(B^+\to K^*\gamma)$ & $3.43 \pm 0.84$ && $4.21 \pm 0.18$ & \cite{Amhis:2014hma} \\
$10^{5}\times\text{BR}(B^0\to K^*\gamma)$ & $3.48 \pm 0.81$ && $4.33 \pm 0.15$ & \cite{Amhis:2014hma} \\
$10^{5}\times\overline{\text{BR}}(B_s\to \phi\gamma)$ & $4.31 \pm 0.86$ && $3.5 \pm 0.4$ & \cite{Aaij:2012ita,Dutta:2014sxo} \\
$S(B^0\to K^*\gamma)$ & $-0.023\pm 0.015$ && $-0.16\pm0.22$ & \cite{Amhis:2014hma} \\
$A_\text{CP}(B^0\to K^*\gamma)$ & $0.003\pm 0.001$ && $-0.002\pm0.015$ & \cite{Amhis:2014hma} \\
$A_{\Delta\Gamma}(B_s\to \phi\gamma)$ & $0.031\pm0.021$ && $-1.0\pm0.5$ & \cite{Aaij:2016ofv}\\
$\langle P_1 \rangle(B^0\to K^*e^+e^-)_{[0.002,1.12]}$ & $0.04\pm0.02$ && $-0.23\pm0.24$ & \cite{Aaij:2015dea} \\
$\langle A_T^\text{Im} \rangle(B^0\to K^*e^+e^-)_{[0.002,1.12]}$ & $0.0003\pm0.0002$ && $0.14\pm0.23$ & \cite{Aaij:2015dea} \\
\hline
\end{tabular}
\caption{SM predictions vs.\ experimental world averages of
observables sensitive to the $b\to s\gamma$ transition.}
\label{tab:obs}
\end{table}

\section{Numerical analysis}\label{sec:num}

In our numerical analysis we will primarily use \flavio \cite{flavio} and cross-check all
the results with \HEPfit \cite{HEPfit}.
While all the analysis in this work will be made available as a standard part
of a \flavio release,
the necessary modifications for \HEPfit, tuned to match the numerics generated by \flavio,
can be made available on request.
The details of the \flavio implementation can be found in appendix \ref{app:flavio}.
In the following sections we will look at the SM predictions for the observables
that we have discussed previously. We will also see how these observables can
constrain contributions to $C_7$ and $C_7'$ generated by new physics. As a
separate exercise we try to fit for the values of the form factors assuming
the experimental numbers to be SM signals.

\subsection{Standard Model predictions facing measurements}

The CP and isospin averaged branching ratio of the inclusive $B\to X_s\gamma$
decay has been measured at CLEO~\cite{Chen:2001fja},
Belle~\cite{Abe:2001hk,Limosani:2009qg} and
BaBar~\cite{Aubert:2007my,Lees:2012ym,Lees:2012ufa,Lees:2012wg}.
The HFAG world average for a minimum photon energy of 1.6~GeV 
is shown in table~\ref{tab:obs} and is in excellent agreement with the
SM prediction shown in the same table.

Measurements of the exclusive $B\to K^*\gamma$ branching ratio
have been reported by CLEOII~\cite{Coan:1999kh}, Belle~\cite{Nakao:2004th} and
BaBar~\cite{Aubert:2009ak}.
The current HFAG world averages for the charged and neutral modes are shown
in table~\ref{tab:obs} along with our SM predictions\footnote{%
By including the branching ratios of both the charged and neutral mode,
we implicitly include the isospin asymmetry as a constraint, which is
generated by the interference of weak annihilation and $C_7$.
Note however that the current HFAG averages do not take into account the
different treatment of the $B^+/B^0$ production asymmetry at $B$ factories,
which leads to a bias on the asymmetry \cite{Jung:2015yma}.}.

The first measurement of the (time-integrated) branching ratio of
$B_s\to\phi\gamma$ has been performed by Belle \cite{Wicht:2007ni}.
LHCb has presented a measurement of the ratio of branching ratios of
$B_s\to\phi\gamma$ and $B^0\to K^*\gamma$ \cite{Aaij:2012ita}, which can be
converted to a measurement of $\text{BR}(B_s\to\phi\gamma)$ by using the
world average of $\text{BR}(B^0\to K^*\gamma)$ from other experiments.
Again,  table~\ref{tab:obs} compares our SM prediction
with the HFAG world average obtained in this way. 

Comparing the SM predictions with the measurements of the three exclusive
branching ratios, one notices that, although they agree at the level of
$1\sigma$, the SM predictions tend to be on the low side for $B\to K^*\gamma$
and on the high side for $B_s\to\phi\gamma$. Such a pattern -- if it were 
significant -- could \textit{not} be explained by NP, which would affect both decays
in the same way. Likewise, it is unlikely to be explained by a hadronic effect
encoded in $\Delta C_7^{(\prime)}$ since it would either affect
all three branching ratios in a similar way (for spectator-independent effects)
or one would expect a larger difference between the $B^+$ and $B^0$ decay than
between the $B^0$ and $B_s$ decay (for effects depending on the spectator
charge). Consequently, if this pattern persists with more precise measurements,
it would point towards a higher value for the $B\to K^*$ form factor $T_1(0)$
and a lower value for the corresponding $B_s\to \phi$ form factor than in
eqs. \eqref{eq:T1Ks} and \eqref{eq:T1phi}, respectively.

\begin{figure}[tbp]
\centering
\includegraphics[width=0.75\textwidth]{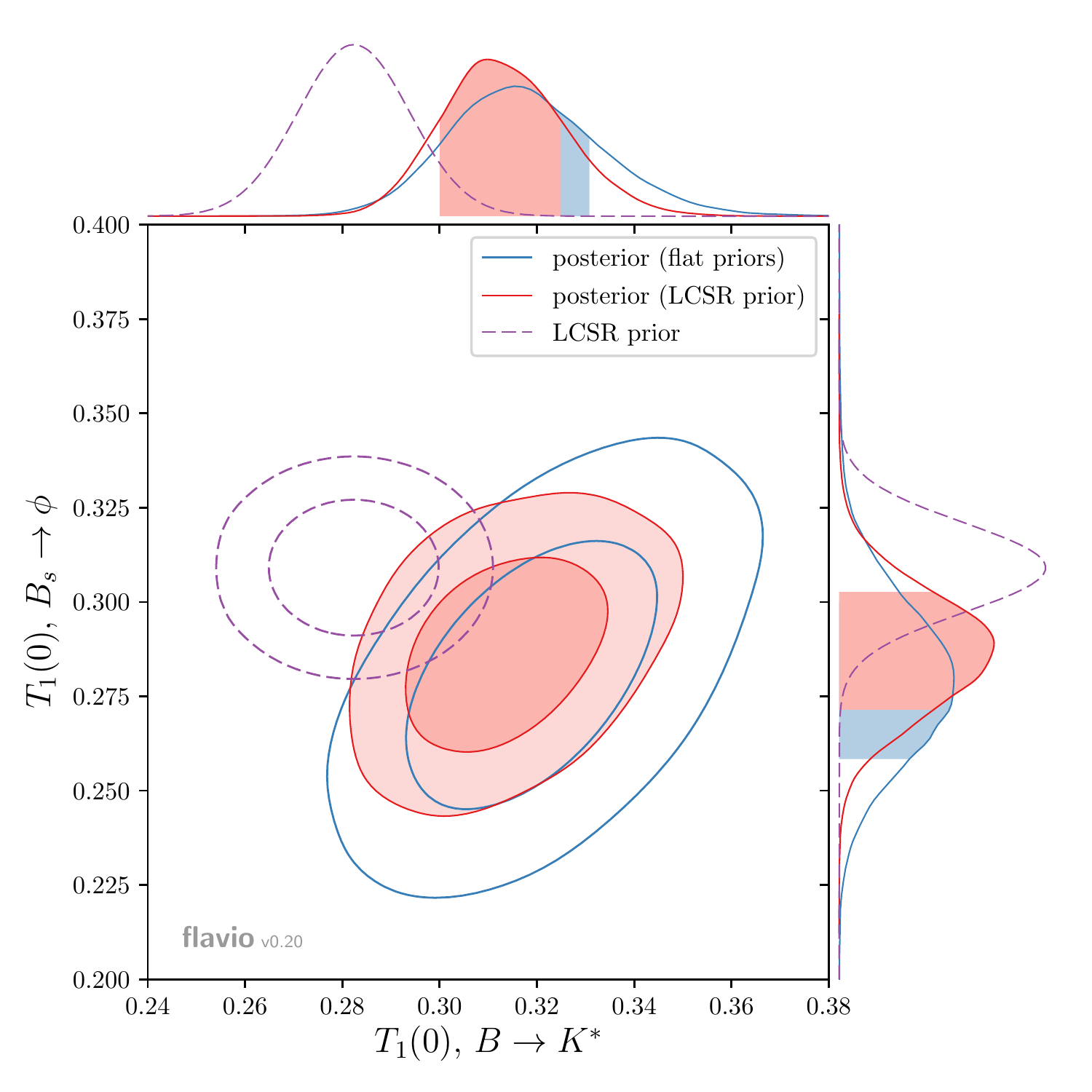}
\caption{Posterior probability distribution of the tensor form factors $T_1(0)$
in $B\to K^*$ and $B_s\to \phi$ transitions from two Bayesian fits to the
three exclusive radiative branching ratios, assuming flat priors
(blue, empty contours) or the LCSR priors, \eqref{eq:T1} (red, filled contours).
The dashed contours show the LCSR priors, \eqref{eq:T1}, for comparison.
Contours correspond to $68.3$ and $95.5\%$ posterior probability.
The curves along the top and right axes show the one-dimensional marginal
(posterior or prior) distributions for the individual form factors; the
shaded bands correspond to $68.3\%$ marginal probability.}
\label{fig:ff_bayes}
\end{figure}

To quantify this statement, we have performed two Bayesian fits of the relevant
hadronic quantities to the measured exclusive branching ratios 
$\text{BR}(B^0\to K^*\gamma)$, $\text{BR}(B^+\to K^*\gamma)$, and
$\text{BR}(B_s\to\phi\gamma)$ using a Markov Chain Monte Carlo
(with \flavio\ and \texttt{emcee} \cite{ForemanMackey:2012ig}),
assuming the validity of the SM. For the first fit, we assumed 
flat priors for the form factors $T_1(0)$, for the second fit we used
the LCSR priors, \eqref{eq:T1}.
The posterior distribution for the form factors for both fits are shown
in fig.~\ref{fig:ff_bayes}, compared to the LCSR prior.
For the fit with \textit{flat priors},
we find the following mean and variance of the form factors,
\begin{align}
T_1(0) &= 0.316 ^{+0.016} _{-0.015} &&\text{for } B\to K^*\gamma
\,,\\
T_1(0) &= 0.280 ^{+0.020} _{-0.022} &&\text{for } B_s\to \phi\gamma
\,.
\label{eq:T1num}
\end{align}
These can be seen as fit predictions given the data, if the SM holds.

The mixing-induced CP asymmetry $S_{K^*\gamma}$
has been measured\footnote{%
Belle and BaBar have also measured 
$S_{K_S\pi^0\gamma}$ including the resonant regions of
$K^*\gamma$. The HFAG average for this quantity is
$S_{K_S\pi^0\gamma}=-0.15\pm0.20$. We do not use it in
our numerics.
} by both Belle and BaBar
\cite{Aubert:2008gy,Ushiroda:2006fi}, the HFAG average is shown in
table~\ref{tab:obs}. While these measurements are still far from the small
SM prediction, Belle-II is expected to measure $S_{K^*\gamma}\sim3\%$ with
50 $\text{ab}^{-1}$ of data~\cite{Aushev:2010bq}.

The direct CP asymmetry $A_\text{CP}(B^0\to K^*\gamma)$ has been measured on
the one hand in the time-dependent CP asymmetry in $B^0\to K^*(\to K_S\pi^0)\gamma$
by BaBar and Belle. On the other hand, as discussed in section~\ref{sec:ex},
it coincides with the \textit{time-integrated} CP asymmetry, which has been
measured much more precisely. The HFAG world average shown
in table~\ref{tab:obs} includes measurements from BaBar \cite{Aubert:2009ak}
and LHCb \cite{Aaij:2012ita}.

Angular observables of the decay $B^0\to K^*e^+e^-$ at low $q^2$ have been
measured only by LHCb so far \cite{Aaij:2015dea}.
The same holds for the mass-eigenstate rate asymmetry $A_{\Delta\Gamma}$ in 
$B_s\to \phi\gamma$, that has been measured very recently for the first
time \cite{Aaij:2016ofv}, with a large negative central value but still large
uncertainty. While this measurement is two standard deviations away from
the SM expectation, we stress that values below $-1$ are actually unphysical
as seen from \eqref{eq:ADeltaGamma}.

For all the observables that vanish in the limit of purely left-handed
photon polarization, i.e. 
$S(B^0\to K^*\gamma)$,
$A_{\Delta\Gamma}(B_s\to \phi\gamma)$,
and the binned angular observables 
$\langle P_1 \rangle$ and $\langle A_T^\text{Im} \rangle$
in $B^0\to K^*e^+e^-$, the measurements are still far away from the small SM
predictions. From eqs.
\eqref{eq:SKsgamma}, \eqref{eq:ADeltaGamma},  \eqref{eq:P1}, and  \eqref{eq:ATIm},
it can be seen that the SM predictions are proportional to
$\mathcal C_7'=(m_s/m_b)C_7 + \Delta C_7$ and thus depend crucially on the
assumption made on the hadronic contribution $\Delta C_7$, that we have
discussed in section~\ref{sec:hadronic}.

\subsection{Constraints on Wilson coefficients}

To constrain the parameter space of NP models, it is crucial to know the 
allowed values of the NP contributions to the Wilson coefficients $C_7$
and $C_7'$. In this section, we discuss the constraints separately for
the real and imaginary parts of $C_7$ as well as for $C_7'$.

\subsubsection*{Real part of $C_7$}

\begin{figure}[tbp]
\centering
\includegraphics[height=9cm]{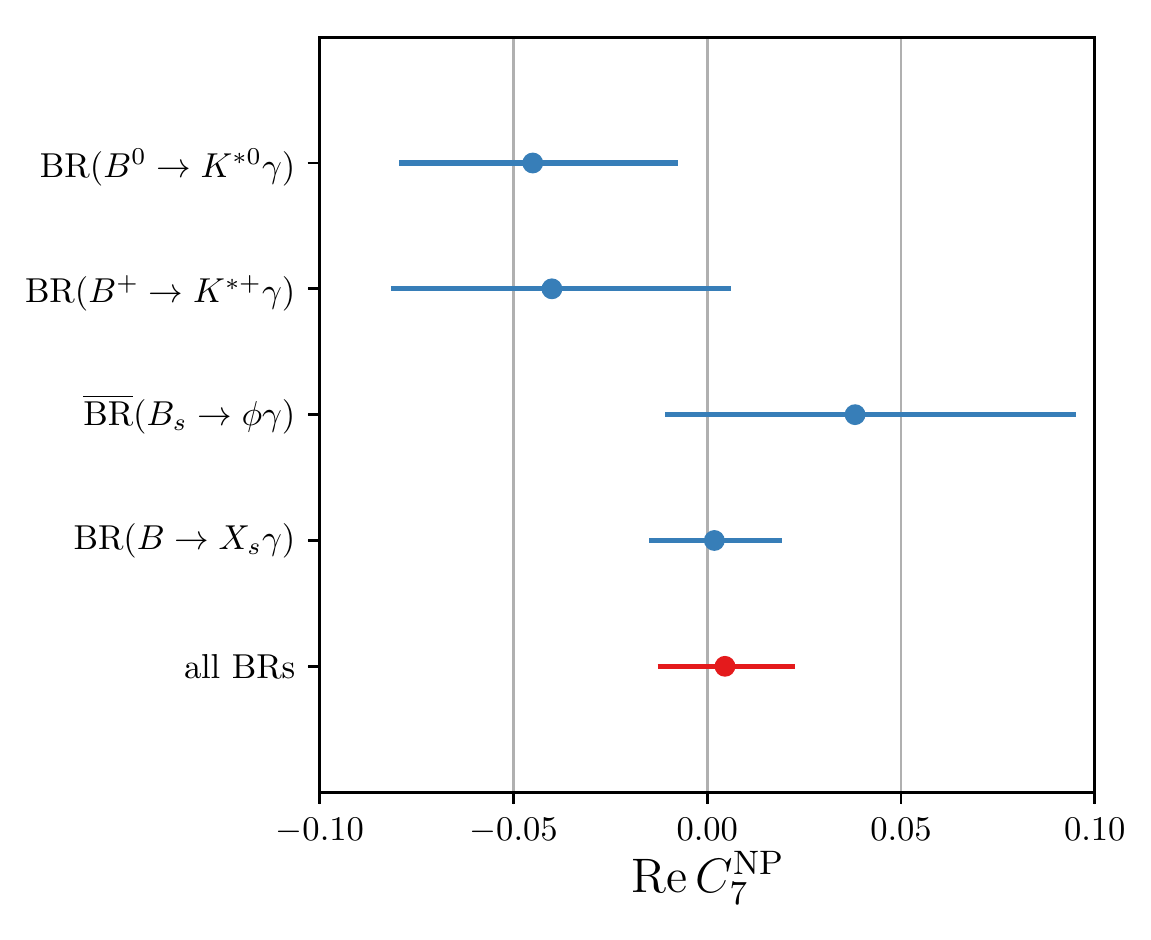}
\caption{Constraint on NP contributions to the real part of the Wilson
coefficient $C_7$ from exclusive and inclusive branching ratios
as well as combined constraint from these branching ratios.}
\label{fig:c7-constraints}
\end{figure}

In scenarios where $C_7'$ does not receive NP contributions
and the contribution to $C_7$ is aligned in phase with the SM, the only
observables sensitive to NP among the ones we have discussed are the branching
ratios. 
Figure~\ref{fig:c7-constraints} shows the constraints on NP contributions to the
real part of $C_7$ from the individual exclusive and the inclusive branching
ratio, as well as the global constraint. The latter is of course dominated
by the inclusive branching ratio, which has the smallest theory uncertainty.
We find the following best-fit regions for the NP contribution\footnote{%
With the one-dimensional 1 and $2\sigma$ regions, we refer to regions where
the logarithm of the likelihood is within 1 and 4 of the best-fit value, respectively.},
\begin{equation}
\text{Re}\,C_7^\text{NP}(\mu_b) \in
\begin{cases}
[-0.018, 0.012] & \text{@ }1\sigma,
\\
[-0.032, 0.027] & \text{@ }2\sigma,
\label{eq:ReC7num}
\end{cases}
\end{equation}
where $\mu_b=4.8~\text{GeV}$.

\subsubsection*{Imaginary part of $C_7$}

As discussed in sec.~\ref{sec:ex}, the only stringent constraint on the imaginary
part of $C_7^\text{NP}$ is expected to come from $A_\text{CP}(B\to K^*\gamma)$.
Using the experimental measurement in table~\ref{tab:obs}, we find
\begin{equation}
\text{Im}\, C_7^\text{NP}(\mu_b) \in [-0.064, 0.094] \times \left[\frac{-0.027}{\text{Im}\, \Delta C_7}\right] \qquad \text{@ 95\% C.L.}
\label{eq:ImC7num}
\end{equation}
Using our numerics and theory error estimates detailed in
section~\ref{sec:hadronic}, we find
\begin{equation}
\text{Im}\,\Delta C_7(\mu_b) = -0.027\pm0.016 \qquad \text{for } B^0\to K^*\gamma \,,
\label{eq:DC7num}
\end{equation}
where the central value is dominated by vertex corrections and spectator
scattering (cf.\ table~\ref{tab:deltaC7}) and the uncertainty by our estimate of neglected contributions, including the soft gluon correction to the charm loop.
From \eqref{eq:DC7num} it is clear that an accidental cancellation in the
imaginary part of $\Delta C_7$, that would make $A_\text{CP}$ tiny even in 
the presence of NP in $\text{Im}\, C_7$, is not entirely excluded. We note
that the estimate of the soft gluon contribution in \eqref{eq:DC7-Zwicky},
that we omitted, would make the constraint even stronger.
In any case, a better
understanding of the hadronic contributions is crucial to better constrain
this Wilson coefficient.

\subsubsection*{Constraints on $C_7'$}

\begin{figure}[tbp]
\centering
\includegraphics[height=6cm]{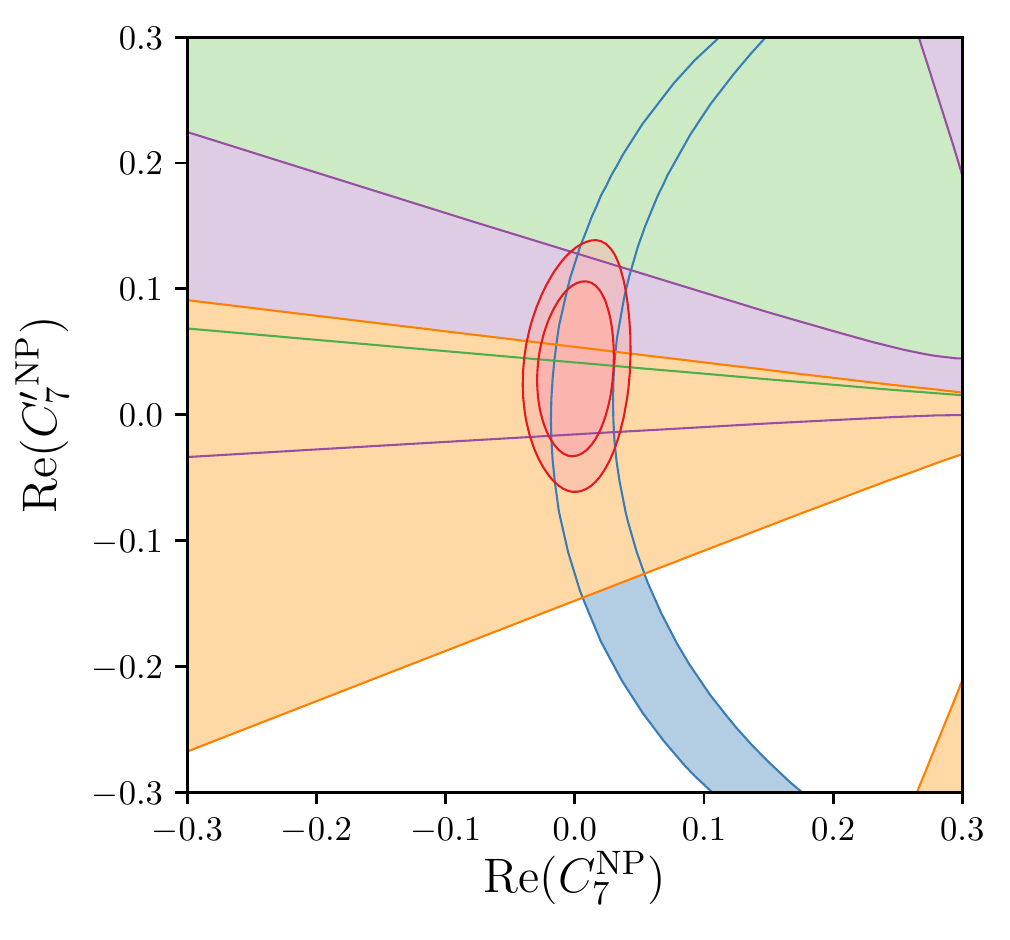}%
\includegraphics[height=6cm]{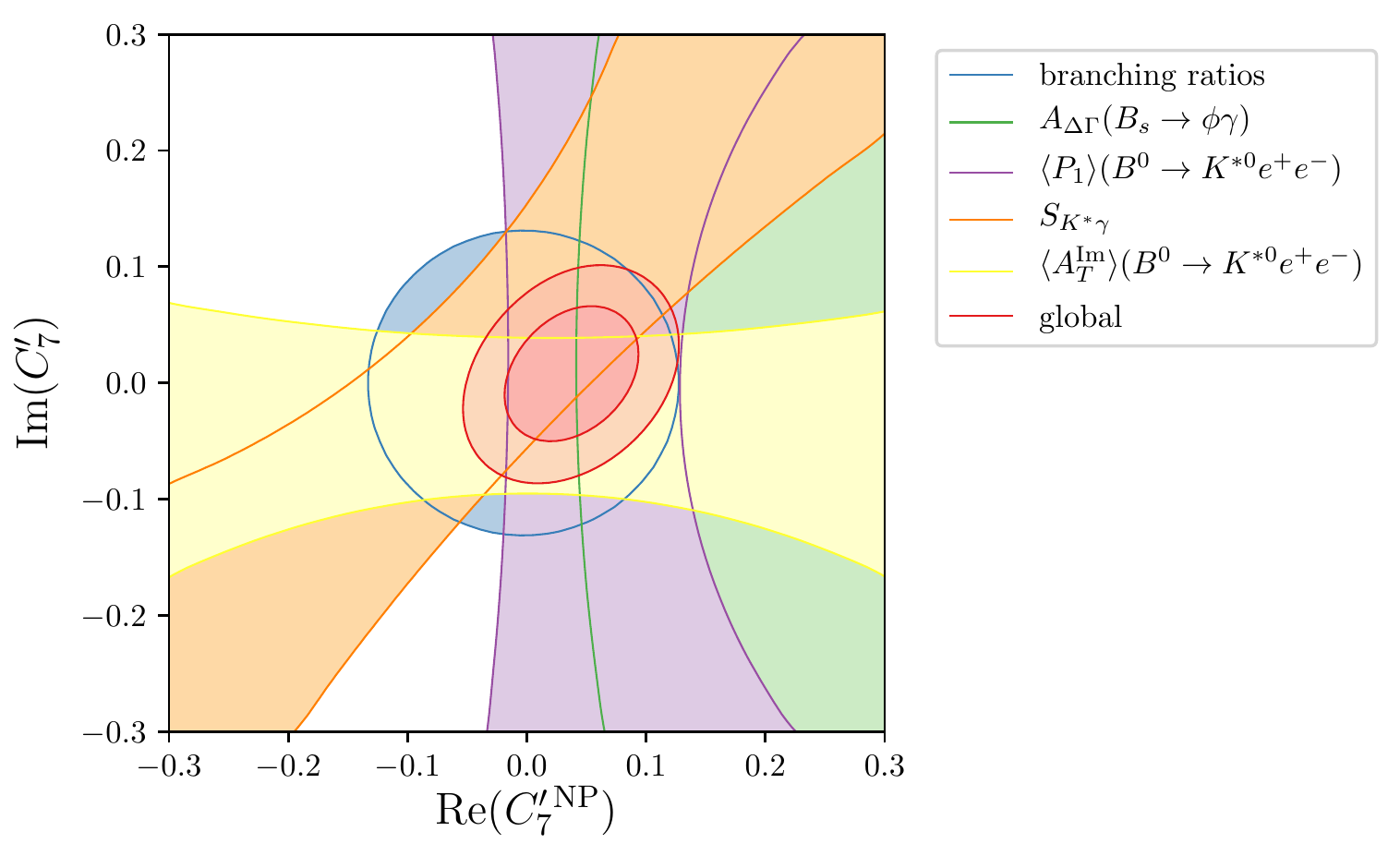}
\caption{Constraints on NP contributions to the Wilson coefficients $C_7$ and
$C_7'$. For the global constraints, 1 and $2\sigma$ contours are shown, while
the individual constraints are shown at $1\sigma$ level.}
\label{fig:c7p-constraints}
\end{figure}

The virtues of the exclusive observables come to play in models predicting a
NP contribution to the ``wrong-chirality'' Wilson coefficient $C_7'$.
In fig.~\ref{fig:c7p-constraints}, we show the constraints in the plane of
NP contributions to
$\text{Re}\,C_7$ vs.\ $\text{Re}\,C_7'$
and
$\text{Re}\,C_7'$ vs.\ $\text{Im}\,C_7'$.
The contours correspond to constant values of $\Delta\chi^2$ with respect to
a best fit point,
obtained by combining (correlated) experimental and theoretical uncertainties\footnote{%
See \cite{Altmannshofer:2014rta} and the documentation of the \texttt{FastFit}
class in \flavio\ for details on the procedure.}.
In each of the plots, we have assumed NP to only affect the two quantities
plotted (e.g., in the first plot, both coefficients are assumed to be real).
In addition to the
global 1 and $2\sigma$ constraints, we also show the $1\sigma$ constraints
from individual exclusive observables as well as from the combination of all
branching ratios.
These plots highlight the complementarity of the exclusive observables:
while the imaginary part of $C_7'$ is constrained by $A_T^\text{Im}$,
the real part is constrained by $A_{\Delta\Gamma}$ and $P_1$, while
$S_{K^*\gamma}$ leads to a constraint in the complex $C_7'$ plane that is
``rotated'' by the $B^0$ mixing phase $2\beta$.
The new measurement of $A_{\Delta\Gamma}$ 
shows a preference for non-zero $\text{Re}\,C_7'$, but given its large
uncertainties, it is not in disagreement
with the measurement of $P_1$.

Since the experimental central value of $A_{\Delta\Gamma}$ is at the border
of the physical domain, we provide best fit values and correlated errors
on the real and imaginary parts of $C_7'$ in a fit without $A_{\Delta\Gamma}$
and in a fit including it, obtained by
approximating the likelihood in the vicinity of the best fit point as a
multivariate Gaussian.
We find
\begin{align}
\begin{pmatrix}
\text{Re}\,C_7^{\prime \, \text{NP}}(\mu_b) \\
\text{Im}\,C_7'(\mu_b)
\end{pmatrix}
&=
\begin{pmatrix}
0.018 \pm 0.037 \\
0.001 \pm 0.037
\end{pmatrix}
,
&\rho &= 0.34
&&\text{(without $A_{\Delta\Gamma}$),}
\\
\begin{pmatrix}
\text{Re}\,C_7^{\prime \, \text{NP}}(\mu_b) \\
\text{Im}\,C_7'(\mu_b)
\end{pmatrix}
&=
\begin{pmatrix}
0.038 \pm 0.035 \\
0.006 \pm 0.036
\end{pmatrix}
,
&\rho &= 0.29
&&\text{(with $A_{\Delta\Gamma}$),}
\end{align}
where $\rho$ are the correlation coefficients.

\section{Conclusions and outlook}\label{sec:conclusions}

The $b\to s\gamma$ transition belongs to the most important probes of NP in the flavour
sector. While the most stringent constraint on new contributions with
left-handed photon helicity comes from the branching ratio of the inclusive decay
$B\to X_s\gamma$, the exclusive radiative decays $B_q\to V\gamma$ and the
semileptonic decays $B_q\to V e^+e^-$ at low $q^2$ are complementary probes
that are sensitive to the photon helicity. In this paper, after having
critically reviewed all the hadronic uncertainties in exclusive radiative and
semi-leptonic decays, we have updated the numerical analysis of new physics in
the Wilson coefficients $C_7$ and $C_7'$ of the electromagnetic dipole
operators, taking into account the recent measurements of the $B\to K^*e^+e^-$
angular distribution and the untagged time-dependent $B_s\to \phi\gamma$ decay
rate by LHCb and emphasizing the role played by the direct CP asymmetry in
$B\to K^*\gamma$.

Our main findings can be summarized as follows.
\begin{itemize}
 \item The inclusive and exclusive branching ratios strongly constrain
 NP contributions to the real part of $C_7$, cf.~\eqref{eq:ReC7num}.
 \item Assuming the SM to hold, the exclusive branching ratios can also be used
 to extract the form factors, cf.~\eqref{eq:T1num}.
 \item The observable most sensitive to the imaginary part of $C_7$ is the
 direct CP asymmetry in $B\to K^*\gamma$, cf.~\eqref{eq:ImC7num}. However, the contribution is proportional
 to the sine of a strong phase that is rather uncertain, adopting our conservative
 error estimates. Improved determinations of this phase would be useful to better
 constrain this Wilson coefficient.
 \item The Wilson coefficient $C_7'$ is constrained by a number of theoretically
 clean\footnote{Given present experimental uncertainties} observables
 with complementary dependence on the Wilson coefficients:
 the mixing-induced CP asymmetry in $B\to K^*\gamma$, the angular observables
 $P_1$ and $A_T^\text{Im}$ in $B\to K^*e^+e^-$ at low $q^2$, and the mass-eigenstate
 rate asymmetry $A_{\Delta\Gamma}$ in $B_s\to\phi\gamma$ measured recently for the first time.
 The new measurement of $A_{\Delta\Gamma}$ shows a slight preference
 for non-standard $C_7'$, but the global fit does not show a significant tension.
\end{itemize}
While we have only presented one- and two-dimensional constraints on Wilson
coefficients, a global Bayesian fit
simultaneously fitting all Wilson coefficients as well as the hadronic
contributions would be interesting to quantify the agreement of different hypothesis
on NP or hadronic contributions given the data.
We leave this exercise to a future update of this work.

In our numerical analysis, we have purely relied on open source codes,
in particular \flavio\footnote{\url{https://flav-io.github.io}} and \HEPfit\footnote{\url{http://hepfit.roma1.infn.it}}. Appendix~\ref{app:flavio}
contains some details on how to modify \flavio\ to study the impact of different
parameter choices.
These public codes can also play an instrumental role in improving the
constraints on new physics in $C_7^{(\prime)}$ by future measurements, e.g.\
by LHCb or Belle-II. In addition to improved measurements of the $B\to K^*\gamma$
and $B_s\to\phi\gamma$ branching ratios, this includes in particular
\begin{itemize}
 \item measurements of the radiative baryonic decays $\Lambda_b\to\Lambda^{(*)}\gamma$
 \cite{Mannel:1997xy,Hiller:2001zj},
 \item a more precise measurement of the time-dependent CP asymmetry in
 $B^0\to K^*\gamma$ to zero in on $S_{K^*\gamma}$,
 \item Improved measurements of the $B\to K^*e^+e^-$ angular analysis at very low
 $q^2$ and the analogous measurement in $B_s\to \phi e^+ e^-$.
\end{itemize}
On the theory side, the main limiting factor in exclusive decays is the form
factor uncertainty, impeding the exploitation of the precise branching ratio
measurements. A moderate improvement might be possible in the future from
extrapolations of improved lattice calculations of the form factors at high
$q^2$, in particular in the $B_s\to\phi$ case, while the $B\to K^*$ form factors
are more challenging due to the large $K^*$ width \cite{Agadjanov:2016fbd}.
Concerning angular observables, mixing-induced CP asymmetries and
$A_{\Delta\Gamma}$, these observables are instead virtually unaffected by the form factor
uncertainty. Their uncertainty is dominated by poorly known contributions to
the hadronic quantities $\Delta C_7^{(\prime)}$ that would profit from more
precise estimates in the future. Nevertheless, given their smallness within the
SM, these uncertainties will be subdominant compared to the experimental
uncertainties for the next few years, unless a sizable deviation from the SM
expectation is observed.

\section*{Acknowledgements}

D.~S.\ thanks Wolfang Altmannshofer, Christoph Bobeth, Martino Borsato, Martin Jung, 
Alexander Khodjamirian,
Mikołaj Misiak, Marie-Hélène Schune
and Roman Zwicky
for useful discussions
and Thomas Blake for an important bug report on flavio.
The work of D.~S.\ is supported by the DFG cluster of excellence ``Origin and
Structure of the Universe''. A.~P. would like to thank Emi Kou for discussions related to this work. A.~P. is supported by the European Research Council under the European Union's Seventh Framework Programme (FP/2007-2013)/ERC Grant Agreement n.~279972 `NPFlavour'.

\appendix

\section{Reproducing numerics with \text{flavio}}\label{app:flavio}

The Standard Model predictions and plots in this paper have been obtained
with the open source Python package \flavio, version 0.20.
For usage documentation and details on this code, see its web site\footnote{%
\url{https://flav-io.github.io}}.
The central values
and uncertainties of all parameters discussed in this paper correspond to the
default values of this version, with the exception of the $B\to V$ form factors,
that we take from a LCSR calculation, while \flavio by default uses a combined
fit to lattice and LCSR results. The LCSR form factors can be loaded as
default in any script or session, after invoking \verb~import flavio~,
with the commands
\begin{verbatim}
from flavio.physics.bdecays.formfactors.b_v import bsz_parameters
bsz_parameters.bsz_load_v2_lcsr(flavio.default_parameters)
\end{verbatim}
The SM central values and uncertainties of the radiative decay observables can
be computed with the commands
\begin{verbatim}
flavio.sm_prediction(<obs>)
flavio.sm_uncertainty(<obs>)
\end{verbatim}
where \verb~<obs>~ has to be replaced by
\begin{itemize}
 \item \verb~'BR(B->Xsgamma)'~ for $\text{BR}(B\to X_s\gamma)_{E_\gamma>1.6\,\text{GeV}}$,
 \item \verb~'BR(B0->K*gamma)'~ for $\text{BR}(B^0\to K^{*0}\gamma)$,
 \item \verb~'BR(B+->K*gamma)'~ for $\text{BR}(B^+\to K^{*+}\gamma)$,
 \item \verb~'BR(Bs->phigamma)'~ for $\overline{\text{BR}}(B_s\to \phi\gamma)$,
 \item \verb~'S_K*gamma'~ for $S_{K^*\gamma}$,
 \item \verb~'ADeltaGamma(Bs->phigamma)'~ for $A_{\Delta\Gamma}(B_s\to \phi\gamma)$.
\end{itemize}
For the $B\to K^*e^+e^-$ observables, the analogous commands read
\begin{verbatim}
flavio.sm_prediction(<obs>, q2min=0.002, q2max=1.12)
flavio.sm_uncertainty(<obs>, q2min=0.002, q2max=1.12)
\end{verbatim}
where \verb~<obs>~ is
\begin{itemize}
 \item \verb~'<P1>(B0->K*ee)'~ for $\langle P_1 \rangle$,
 \item \verb~'<ATIm>(B0->K*ee)'~ for $\langle A_T^\text{Im} \rangle$.
\end{itemize}

The easiest way to study the impact of different parameter or theory uncertainty
choices is to modify the default parameter values. For instance, to set the
$B\to K^*$ form factor $T_1(0)$ to $0.3\pm0.1$, use
\begin{verbatim}
flavio.default_parameters.set_constraint('B->K* BSZ a0_T1', '0.3 +- 0.1')
\end{verbatim}
The other most relevant parameters for exclusive radiative decays are
\begin{itemize}
 \item \verb~'Bs->phi BSZ a0_T1'~ -- $B_s\to\phi$ form factor $T_1(0)$
 \item \verb~'B0->K*0 deltaC7p a_+ Re'~ -- $\text{Re}(\Delta C_7')$ in $B^0\to K^{*0}\gamma$
 \item \verb~'B0->K*0 deltaC7 a_- Re'~ -- $\text{Re}(\Delta C_7)$ in $B^0\to K^{*0}\gamma$
\begin{itemize}
 \item analogously for \verb~'B+->K*+'~ and \verb~'Bs->phi'~
 \item analogously for the imaginary parts of $\Delta C_7^{(\prime)}$ using
 \verb~'Re'~ $\to$ \verb~'Im'~.
\end{itemize}
\end{itemize}

A Jupyter notebook to reproduce the plots in fig.~\ref{fig:c7p-constraints}
can be found in a public repository: \cite{david_straub_2017_375593}.

\bibliographystyle{JHEP}
\bibliography{bib}

\end{document}